\documentclass[global,twocolumn]{svjour}

\usepackage{graphicx}

\usepackage{units}

\journalname{}

\newcommand{\Yb}{Yb${}^+$}
\newcommand{\tsub}[1]{\textnormal{\tiny{#1}}}

\begin{document}

\title{Simple vibration insensitive cavity for laser stabilization at the $10^{-16}$ level}

\author{J. Keller\inst{1} \and S. Ignatovich\inst{2} \and S. A. Webster\inst{3} \and T. E. Mehlst\"aubler\inst{1}}

\institute{Physikalisch-Technische Bundesanstalt, Bundesallee 100, 38116 Braunschweig, Germany,\\\email{tanja.mehlstaeubler@ptb.de} \and Institute of Laser Physics, 630090 Novosibirsk, Russia \and M Squared Lasers Ltd, 1 Kelvin Campus, Maryhill Road, Glasgow, G20 0SP, Scotland}

\date{Received: date / Revised version: date}

\maketitle

\begin{abstract}
We present the design and realization of two reference cavities for ultra-stable lasers addressing narrow transitions in mixed-species (${}^{115}$In${}^+$~/~${}^{172}$Yb${}^+$) Coulomb crystals. With a simple setup, we achieve a fractional frequency instability close to the thermal noise limit of a $\unit[12]{cm}$ long cavity, reaching $\sigma_y=4.7\times10^{-16}$ at $\unit[10]{s}$ with a linear drift of $\unit[53]{mHz/s}$. We discuss the individual instability contributions and show that in a setup with a lower thermal noise floor and vibration sensitivity, an instability of $1\times10^{-16}$ can be reached. To achieve this, we implement a vibration insensitive design for a $\unit[30]{cm}$ long cavity mounted horizontally and conduct first tests that show a sensitivity of $\unit[1.8\times10^{-11}]{ms^{-2}}$ to vertical accelerations. This is about a factor of $20$ less than the value observed for the short cavity. Mechanical tolerances and ways to further reduce the sensitivity are discussed.
\end{abstract}

\section{Introduction}

Optical clocks based on single trapped ions are promising candidates for accurate frequency standards, with fractional frequency inaccuracies surpassing $10^{-17}$ \cite{Chou2010}. However, the quantum projection noise (QPN) limited interrogation \cite{Itano1993} of a single particle requires long intervals between frequency corrections applied to the clock laser, thus limiting the achievable frequency instability and setting high demands on the stabilization of the local oscillator.
In order to overcome these limitations, our approach \cite{Herschbach2011,Pyka2012} replaces the time average with an ensemble average by using linear Coulomb crystals containing multiple ($N=10\ldots100$) ${}^{115}$In${}^+$ reference ions sympathetically cooled by ${}^{172}$Yb${}^+$ ions. For a given averaging time, the QPN limited frequency instability of an uncorrelated ensemble is thereby reduced by a factor of $1/\sqrt{N}$. In a clock based on the ${}^1\textnormal{S}_0\leftrightarrow {}^3\textnormal{P}_0$ transition in ${}^{115}$In${}^+$ (natural linewidth $\Gamma=2\pi\times\unit[0.8]{Hz}$), the QPN limit is at a fractional frequency instability of $\sigma_y=7\times10^{-17}/\sqrt{\tau}$ for $N=100$ (assuming an ideal local oscillator and an interrogation period of $T=1/\Gamma\approx\unit[200]{ms}$) \cite{Peik2006}. Clock operation at this limit imposes two requirements on the short-term instability of the spectroscopy laser: interrogations need to be Fourier-limited (i.e. $\sigma_y(T)\ll4\times10^{-15}$), and $\sigma_y$ must be at a level of $1\times10^{-16}$ for averaging times of about $2T\approx\unit[400]{ms}$, after which it can be stabilized to the atomic transition.
In a single-ion system with the same instability, the stabilization time constant around $\unit[100]{s}$ would require local oscillator instabilities on the order of $10^{-17}$ at that time to achieve the same long-term instability.

Fabry-P\'erot cavities made from ultra-low thermal expansion materials such as ULE$\textsuperscript{\textregistered}$ have proven to be a suitable reference for laser frequency stabilization at these timescales \cite{Young1999}. At the required level of frequency instability, the dominant contributions, affected by the choice of geometry and materials, are deformations due to environmental vibrations and Brownian motion of the cavity constituents (thermal noise) \cite{Numata2004}. Sensitivity to vibrations can be suppressed by using mounting geometries that exploit symmetries \cite{Nazarova2006,Ludlow2007,Webster2008}. The effect of thermal noise can be reduced by using materials with higher mechanical quality (Q) factors such as fused silica (FS) mirror substrates \cite{Millo2009,Dawkins2010}. Several approaches for further reductions have recently brought instabilities $\leq1\times10^{-16}$ into reach. These include the use of even higher-Q materials such as silicon while operating the cavity at cryogenic temperatures \cite{Kessler2012}, beam geometries that cover larger fractions of the mirror surfaces \cite{DAmbrosio2003,Mours2006,Amairi2012}, novel coating materials \cite{Cole2013}, and increased spacer lengths \cite{Amairi2012,Jiang2011,Nicholson2012}.

In this paper, we present two different cavities used to stabilize lasers for precision spectroscopy of mixed In${}^+$~/~Yb${}^+$ Coulomb crystals. The first is used to address the ${}^2$S${}_{1/2} \leftrightarrow {}^2$D${}_{5/2}$ transition in ${}^{172}$Yb${}^+$ with a natural linewidth of $2\pi\times\unit[23]{Hz}$ to perform sideband cooling and studies of the crystal dynamics. A simple setup, consisting of FS mirrors on a $\unit[12]{cm}$ long ULE spacer without a special vibration insensitive design, allowed us to achieve a fractional frequency instability of $4.7\times10^{-16}$ in $\unit[10]{s}$. In order to reach an instability of $1\times10^{-16}$ for the clock laser interrogating the In${}^+$ ions, the second cavity is made of a $\unit[30]{cm}$ long spacer with a vibration insensitive mounting optimized using finite element method (FEM) calculations. Experimentally determined vibration sensitivities are in good agreement with these calculations.

Section \ref{12cmcavity} introduces the simple $\unit[12]{cm}$ cavity setup and the individual contributions to the frequency instability. The instability of a laser stabilized to this cavity, determined by comparison with two other stable lasers, is shown to be consistent with these results. Section \ref{30cmcavity} describes the design of the $\unit[30]{cm}$ clock laser cavity. FEM calculations of the vibration sensitivity  are presented and compared to values that are determined experimentally using the $\unit[12]{cm}$ cavity as a reference.

\section{Stability limit of a simple ULE cavity setup with fused silica mirrors}
\label{12cmcavity}
The laser system described in this section consists of an extended cavity diode laser (ECDL) operating at \unit[822]{nm}, which is frequency doubled and used to address the \unit[23]{Hz} wide ${}^2\textnormal{S}_{1/2}\leftrightarrow {}^2\textnormal{D}_{5/2}$ transition in \Yb. It is stabilized to a cavity which consists of a cylindrical spacer made of ultra-low expansion glass (Corning ULE 7973) with a length of \unit[12]{cm} and a diameter of \unit[6]{cm}. One plane mirror and one curved mirror with a radius of $\unit[1]{m}$ are used, resulting in spot sizes of $\unit[293]{\mu m}$ and $\unit[313]{\mu m}$, respectively. The mirror substrates are made from fused silica in order to reduce the thermal noise floor, which is expected to be at a fractional frequency instability of $\sigma_y=2.9\times10^{-16}$ according to \cite{Numata2004}.

\begin{figure}
\centerline{\includegraphics[width=.5\textwidth]{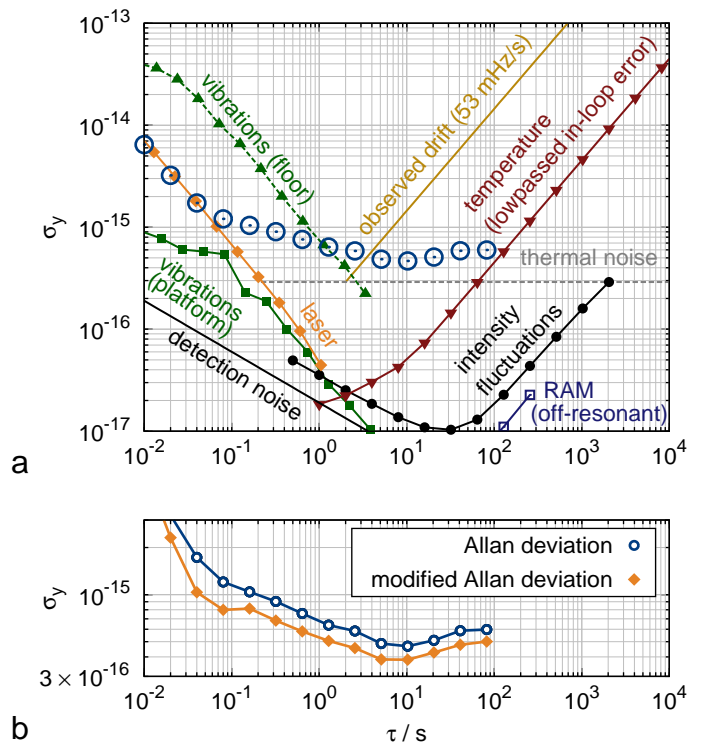}}
\caption{(\textbf{a}) Summary of instability contributions of an ECDL stabilized to a $\unit[12]{cm}$ long ULE cavity. Instabilities are given as fractional frequency Allan deviations. The blue open circles show the resulting instability of the laser (after removing the linear drift) determined in a three-cornered-hat measurement with two other stable lasers \cite{Kessler2012,Haefner}. (\textbf{b}) The modified Allan deviation reveals the peak at $\unit[0.2]{s}$ due to additional vibrations transmitted by a cable.}
\label{instability_summary}
\end{figure}

In the following, we present our experimental setup and analyze the major contributions to the achievable frequency instability in such a simple system without vibration insensitive design.

\subsection{Environmental vibrations}
\label{vibrations}

\begin{figure}
\centerline{\includegraphics[width=.5\textwidth]{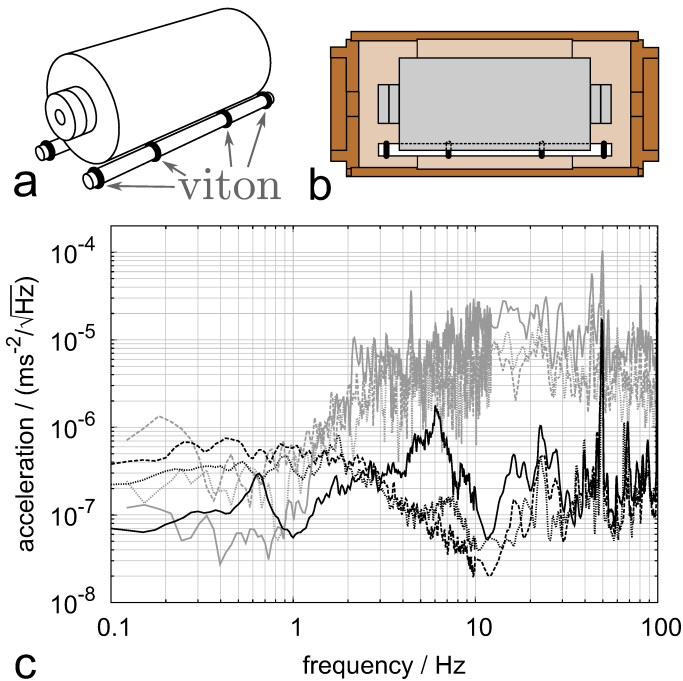}}
\caption{(\textbf{a}) The spacer is supported by four viton rings held in place by fused silica rods. (\textbf{b}) These rods rest on four additional viton rings inside a cylindrical copper heat shield with varying inner diameter. Their transverse distance is fixed by an aluminium clamp (not depicted). (\textbf{c}) Acceleration spectra due to environmental vibrations. The grey curves were recorded on the floor of the laboratory, black curves show the suppressed vibrations on the passive isolation platform supporting the cavity. The solid lines correspond to the vertical, dashed and dotted lines to the horizontal directions. The peak at \unit[50]{Hz} is due to electronic noise.}
\label{measured_vibrations}
\end{figure}

The spacer design is not optimized for vibration insensitivity; the support consists of four viton rings approximately at the longitudinal Airy points of the spacer (see Fig. \ref{measured_vibrations}a,b). An estimate taking only deformation due to the Poisson effect into account (assuming a density of $\unit[2.2\times 10^3]{kg} / \unit{m^3}$, Young's modulus of $\unit[6.8\times 10^{10}]{Pa}$, and Poisson ratio of $0.17$ \cite{ULEdatasheet}) results in a sensitivity of the frequency to vertical accelerations of $\unit[2.8\times10^{-10}]{/(ms^{-2})}$. Since the mounting is symmetric in the other dimensions, lower values are expected for the sensitivities to horizontal vibrations. Experimentally, we observed sensitivities of $\unit[(3.5\pm0.2)\times 10^{-10}]{/(ms^{-2})}$ (vertical accelerations) and below $\unit[10^{-10}]{/(ms^{-2})}$ (horizontal). To reduce the influence of vibrations on the frequency stability, the cavity is set up on a passive vibration isolation platform (Minus-K Technology 150BM-1). Additional acoustic isolation is achieved by encasing the whole setup in a wooden box lined with sound absorbing poly\-urethane mats (BaryCell 1440). Figure \ref{measured_vibrations}c shows the acceleration spectra measured on the laboratory floor as well as the suppressed accelerations on the platform. The contribution to frequency instability is obtained by multiplying these with the respective sensitivities; the corresponding Allan deviations are shown in Fig. \ref{instability_summary} as green squares (triangles) for the case with (without) the passive vibration isolation. While this is more than sufficient for the purpose of this laser, it would be close to the maximum acceptable instability of the clock laser.

\subsection{Temperature variations}
\label{temperature_section}
\begin{figure}
\centerline{\includegraphics[width=.5\textwidth]{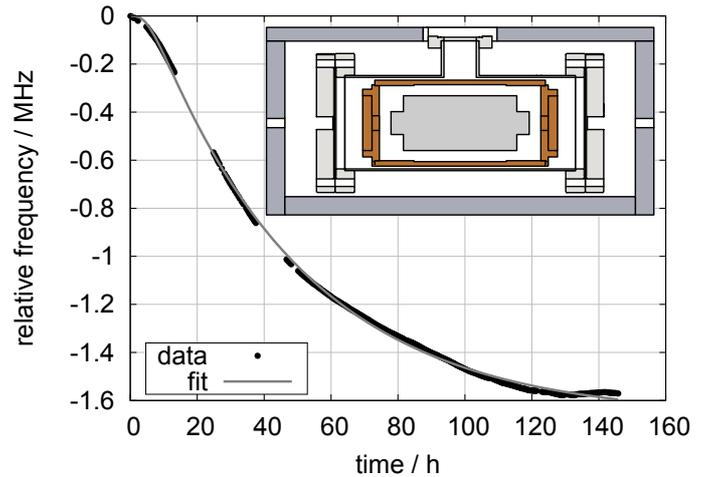}}
\caption{Frequency response to a temperature step (\unit[24.56]{${}^\circ$C} to \unit[24.99]{${}^\circ$C}) measured using a frequency comb referenced to a hydrogen maser. The dots show the experimental data after subtracting the previously observed linear drift, the solid line is a fourth-order low-pass function fitted to the data. The time constants are $\tau_1=\tau_2=\unit[1.8]{h}$ (active stabilization), $\tau_3=\unit[1.9]{h}$ (copper $\leftrightarrow$ cavity), and $\tau_3=\unit[44.4]{h}$ (vacuum chamber $\leftrightarrow$ copper). The inset schematically shows the heat shield configuration.}
\label{temperature_step}
\end{figure}

The spacer material is expected to have a zero crossing of its coefficient of thermal expansion (CTE) around \unit[20]{${}^\circ$C}. However, the CTE mismatch between ULE and FS induces stress at the spacer-mirror interface, which leads to a temperature dependent bulging of the mirrors. This effect, which would reduce the zero crossing temperature of the combined effective CTE to between \unit[-10]{${}^\circ$C} and \unit[0]{${}^\circ$C}, is expected to be suppressed by about a factor of $7$ by ULE rings (\unit[6]{mm} thickness, outer/inner diameter of \unit[25.4]{mm} and \unit[8]{mm}) optically contacted to the back of the mirrors \cite{Legero2010}. This allows the use of a simple temperature stabilization based on resistive heating that keeps the setup slightly above room temperature (at about \unit[25]{${}^\circ$C}), which is still reasonably close to (i.e. less than $\unit[10]{K}$ from) the zero-crossing. The expected temperature sensitivity of $\unit[1.7\times10^{-9}]{\textnormal{K}^{-2}}$ around the zero-crossing \cite{ULEdatasheet} requires temperature fluctuations to be on the order of $\unit[10]{nK}$ on the relevant timescales. As shown schematically in the inset of Fig.~\ref{temperature_step}, two heat shields are used in order to achieve this. To avoid convective heat transfer (and fluctuations of the refractive index), the cavity is inside a vacuum of \unit[$10^{-6}$]{Pa}. Within the vacuum chamber, temperature fluctuations are passively dampened by a copper cylinder with polished surfaces that are gold coated to prevent a decrease in reflectivity due to oxidation. To decouple the cavity mount from the thermal expansion of the copper, the supporting viton rings are held in place by two fused silica rods which are in turn resting on additional viton rings (as shown in Fig.~\ref{temperature_step}). The vacuum chamber is placed inside an aluminium enclosure, which is where the active temperature stabilization is applied. A layer of polystyrene and the acoustic isolation enclosure provide additional decoupling from fluctuations of the room temperature.

To estimate the temperature filtering effect of the copper shield, we model the radiative heat transfer at each stage as a first-order lowpass. Assuming a small deviation from thermal equilibrium at temperature $T$ and equal surface area $A$, the time constant can be estimated as
\begin{equation}
\tau=\frac{\varepsilon_1+\varepsilon_2-\varepsilon_1\varepsilon_2}{4 \sigma \varepsilon_1 \varepsilon_2 A T^3}C\;,
\end{equation}
where $\varepsilon_i$ are the emissivities of the two surfaces, $C$ is the heat capacity of the inner object and $\sigma$ is the Stefan-Boltzmann constant. With this model we estimate a time constant of $1.5$ to $3$ days for the temperature of the copper heat shield and another $\unit[19]{h}$ to $\unit[46]{h}$ before fluctuations reach the cavity. The main uncertainty is due to the unknown emissivities. With an estimated time constant of more than $10$ days, heat conduction through the mount is negligible.\\

For the experimental characterization of the temperature instability, the effect of a temperature step on the resonance frequency has been observed by monitoring the beat note of the laser with a frequency comb referenced to a hydrogen maser. Figure \ref{temperature_step} shows the response after subtracting the linear drift. A fourth-order low-pass function is fitted to the data, with two of the time constants fixed at $\unit[1.8]{h}$, as determined from the error signal of the temperature PI controller. The additional time constants are \unit[44.4(5)]{h} and \unit[1.93(1)]{h}. Using our model, we can attribute the long time constant to the heat transfer between the vacuum chamber ($\varepsilon_\tsub{steel}=0.3$) and the gold-coated copper cylinder ($\varepsilon_\tsub{Cu1}=0.05$). The second time constant corresponds to heat transfer between the copper cylinder and the cavity if emissivities of $\varepsilon_\tsub{Cu2}=0.5$ and $\varepsilon_\tsub{ULE}=0.8$ are assumed. The order of magnitude difference between the two values for the copper cylinder can be explained by the dullness of the inner surface due to imperfections in the electroplating process.

The temperature sensitivity of the frequency is seen to be \unit[$-1.02(5)\times 10^{-8}$]{K${}^{-1}$}, which corresponds to a zero crossing of the composite cavity effective CTE at \unit[19(1)]{${}^\circ$C}. According to \cite{Legero2010}, this implies that the CTE of the spacer material vanishes at about \unit[22]{${}^\circ$C}.

An estimate of the effect of temperature fluctuations on the achievable instability can be made using the error signal of the temperature controller, which has an instability of $\unit[2]{mK}$ in $\unit[1]{s}$. After applying the heat shield lowpass functions, the curve shown in Fig. \ref{instability_summary} (red triangles) is obtained (as this is derived from an in-loop signal, it just serves as a lower bound). It corresponds to a linear drift of \unit[2]{mHz/s}, which is negligible compared to the expected effect due to aging of the spacer material. Experimentally observed frequency variations with respect to a hydrogen maser within a period of $\unit[4\times10^4]{s}$ reveal a linear drift of $\unit[53]{mHz/s}$ and a sinusoidal variation with an amplitude of $\unit[132]{Hz}$ and a period of $\approx\unit[1]{d}$. Higher order frequency fluctuations contribute less than $8\times10^{-15}$ in $\unit[8000]{s}$.

\subsection{Power fluctuations}
Due to the high finesse ($\approx 280,000$), fluctuations of the input power are strongly amplified within the cavity, resulting in temperature changes of the mirror coatings and substrates that shift the resonance frequency. The sensitivity to this effect has been determined to be \unit[\mbox{$2\times10^{-13}$}]{W${}^{-1}$} (intracavity power) by observing the change in frequency when a step of $\unit[\pm5]{\mu W}$ is applied to the input power. A fit to the temporal behaviour of the frequency after the step indicates that two processes with time constants of \unit[0.3]{s} and \unit[5]{s} (observed for both increasing and decreasing power) contribute by an equal amount to this shift. Keeping this contribution below the thermal noise limit at an input power of \unit[25]{$\mu$W} and a coupling efficiency of $0.35$ requires a relative intensity instability on the order of $10^{-4}$, which is realized by an active stabilization using an acousto-optic modulator (AOM). The error signal is derived from the transmitted power to keep fluctuating non-resonant stray light from affecting the stabilization. The contribution to the frequency instability, derived from an out-of-loop measurement of the power instability, is shown in Fig.~\ref{instability_summary} as filled black circles.

\subsection{Laser system and measured instability}
\label{822nmlaser}

\begin{figure}
\centerline{\includegraphics[width=.47\textwidth]{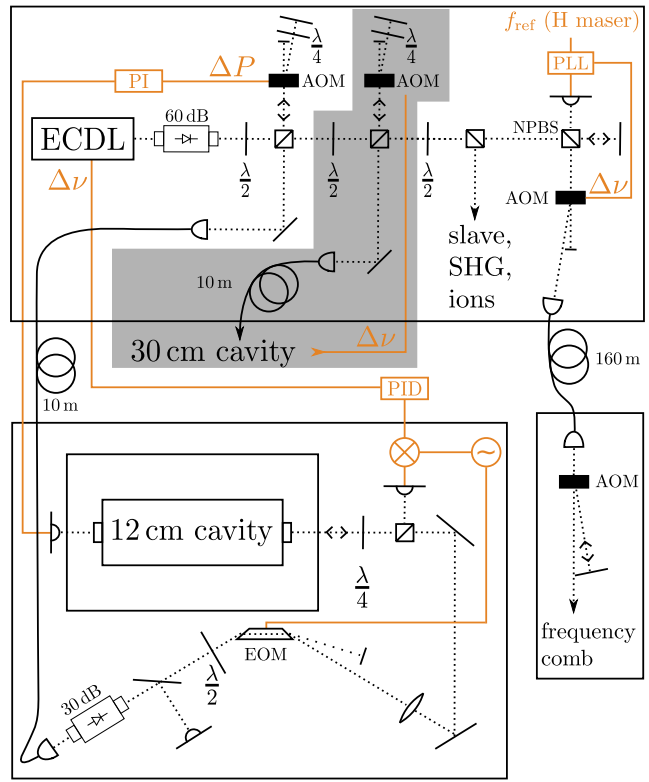}}
\caption{Simplified diagram of the optical setup. All fibres are polarization maintaining, beam splitter cubes are polarizing unless labelled otherwise. The shaded area indicates the temporary addition used for the measurements described in section \ref{30cmcavity}.}
\label{setup}
\end{figure}

The simplified schematic diagram of the complete laser system is shown in Fig. \ref{setup}. A grating stabilized diode laser in Littrow configuration is locked to the cavity by applying feedback to the piezo actuator that adjusts the grating (with a servo bandwidth of $\unit[500]{Hz}$) and to the injection current using a transistor in parallel to the laser diode (bandwidth of $\unit[410]{kHz}$, limited by a resonance in the current modulation of the laser diode at $\unit[1.5]{MHz}$). Unsuppressed Schawlow-Townes noise \cite{Schawlow1958}\\($S_\nu\approx \unit[3\times 10^3]{Hz^2/Hz}$) at higher frequencies is the limiting instability contribution for times below $\unit[0.1]{s}$ as can be seen in Fig. \ref{instability_summary} (orange diamonds). This contribution has been derived from the measured frequency noise spectrum of the free running laser by multiplying it with the gain characteristics of the feedback. It has been experimentally confirmed that this component scales with the inverse square root of the bandwidth (as is expected since the bandwidth determines the averaging time after which the $1\times10^{-13}/\sqrt{\tau}$ white noise contribution is suppressed). Much higher feedback bandwidths have been achieved with diode lasers \cite{Zhao2010} and the intrinsic linewidth could be reduced by extending the external cavity \cite{Littman1978}, but given the purpose of this system, there is no need for further reductions of the short-term instability.

The error signal is generated using the Pound-Drever-Hall method \cite{Drever1983}. The detection noise floor, determined from the power spectral density of the in-loop error signal, is shown as a solid black line in Fig. \ref{instability_summary}. Sidebands at $\unit[32]{MHz}$ are introduced by phase modulation with an electro-optic modulator (EOM) (TimeBase) with a Brewster cut crystal. This geometry spatially separates any residual power in the (non-modulated) ordinary beam from the extraordinary beam and prevents multiple reflections inside the crystal from overlapping, thus reducing residual amplitude modulation (RAM). The RAM component present when the light is off resonance contributes less than $10^{-17}$ for times between $\unit[0.5]{s}$ and $\unit[100]{s}$ (open blue squares in Fig. \ref{instability_summary}). On resonance, an additional component due to parasitic etalons coupled to the cavity could appear that cannot be measured separately.

To characterize the system, the light is transferred to a hydrogen maser referenced frequency comb using a $\unit[160]{m}$ fibre with active phase noise cancellation \cite{Ma1994}. The frequency instability was determined in a three-cornered-hat measurement \cite{Gray1974} using transfer beat notes \cite{Telle2002} with two other stable lasers (a fibre laser at $\unit[1543]{nm}$ stabilized to a cryogenic silicon cavity \cite{Kessler2012} and a diode laser at $\unit[698]{nm}$ stabilized to a ULE cavity \cite{Haefner}). Subtracting the linear drift, a minimum fractional frequency Allan deviation of $4.7\times 10^{-16}$ is reached for an averaging time of $\unit[10]{s}$ as can be seen in Fig. \ref{instability_summary} (open blue circles). For averaging times between $\unit[0.1]{s}$ and $\unit[1]{s}$, there is a discrepancy between the expected contributions and the measured instability. The modified Allan variance shown in Fig. \ref{instability_summary}b identifies this as a peak around $\tau=\unit[0.2]{s}$. Further tests have shown that the source is vibrations transmitted through the power supply cable of the temperature stabilization, which was not present during the measurements shown in  Fig. \ref{measured_vibrations}. With the cable temporarily removed, the laser achieves an instability of $\sigma_y=3.7\times10^{-16}$ in $\unit[4]{s}$. While this is not an issue for the Yb${}^+$ laser system, the cables for the clock laser system have to be chosen more carefully. Cables as soft as RG-58 (or Lemo type 106330 for supplying high voltages) have proven not to be a limitation at this level.

\section{Vibration insensitive design for a \unit[30]{cm} cavity}
\label{30cmcavity}
To achieve QPN limited interrogation with the clock laser addressing the ${}^1$S${}_0\leftrightarrow {}^3$P${}_0$ intercombination line at $\unit[236.5]{nm}$ in an ensemble of $100$ ${}^{115}$In${}^+$ ions, we require a short-term instability of $\sigma_y=1\times10^{-16}$ to be reached within a few hundred milliseconds. Three of the contributions shown in Fig. \ref{instability_summary} need to be addressed in order to achieve this. The first, Schawlow-Townes noise, is expected to be lower for this system, as a frequency quadrupled solid state (Nd:YAG) laser will be used. The second limitation is the thermal noise floor, which can be lowered to $\sigma_y=9.4\times 10^{-17}$ by choosing a spacer length of $\unit[30]{cm}$. The third contribution to be addressed is the influence of vibrations. In section \ref{FEM}, we describe the spacer geometry and mounting we chose in order to minimize the vibration sensitivity and show the results of FEM calculations. Section \ref{sensitivity_measurements} shows the experimental verification of these calculations using the laser described in the previous section.

\begin{figure}
\centerline{\includegraphics[width=.45\textwidth]{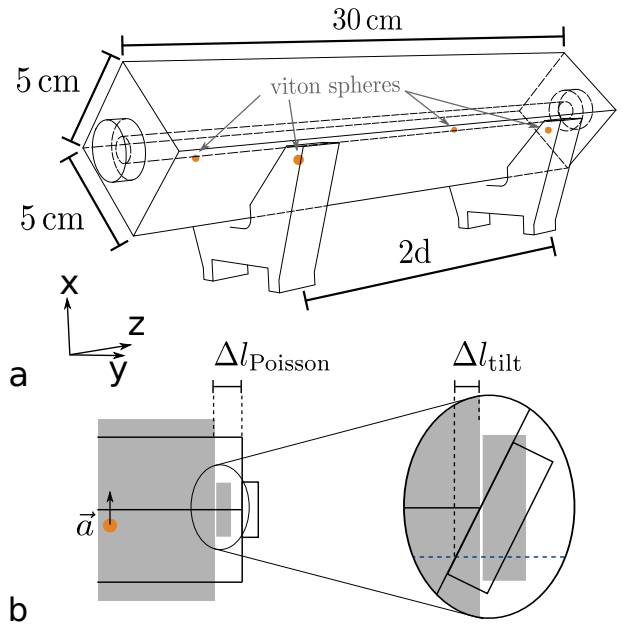}}
\caption{Design for a vibration insensitive \unit[30]{cm} long cavity. (\textbf{a}) Cuboid spacer resting on 4 viton spheres. (\textbf{b}) The optical axis is displaced vertically to allow a cancellation of $\Delta l_\textnormal{\scriptsize{Poisson}}$ by the tilt of the mirrors for vertical accelerations.}
\label{longcavity}
\end{figure}

\subsection{Geometry and FEM calculations}
\label{FEM}
Our design consists of a cuboid resting on viton spheres slightly below its vertical symmetry plane. The supports can be freely moved along the optical axis, as no machining of the spacer is necessary at their positions (see Fig. \ref{longcavity}a). The sensitivity to horizontal vibrations is suppressed by the symmetry of the support. This is not the case for vertical accelerations, where low sensitivities are achieved due to a cancellation of the Poisson effect by the tilt of the mirrors. Because of the $\unit[45]{{}^\circ}$ angle of the support plane, vertical accelerations deform the spacer in both the vertical and transverse direction, making the total Poisson effect independent of the vertical placement of the supports. Assuming the forces to be equally distributed in the respective planes and adding up both components, we estimate the effect as
\begin{equation}
 \left(\frac{\Delta l}{l}\right)_{\textnormal{\scriptsize{Poisson}}}=\frac{\nu\rho W}{\sqrt{2}E}\quad ,
\end{equation}
where $\nu$ is the Poisson ratio, $\rho$ the density, $E$ Young's modulus, and $W$ the width of the spacer. This approximation agrees well with FEM simulations for varied values of $W$. For the experimentally realized case of $W=\unit[5]{cm}$, the expected Poisson effect contribution is $\unit[1.9\times10^{-10}/]{(ms^{-2})}$. If the cavity is supported within the inner \unit[60]{\%} of its length, the mirrors tilt around the vertical symmetry plane. Therefore, a vertical displacement of the optical axis is necessary for the tilt to have an effect on the effective resonator length that compensates for the Poisson effect (as illustrated in Fig. \ref{longcavity}b).

\begin{figure}
\centerline{\includegraphics[width=.45\textwidth]{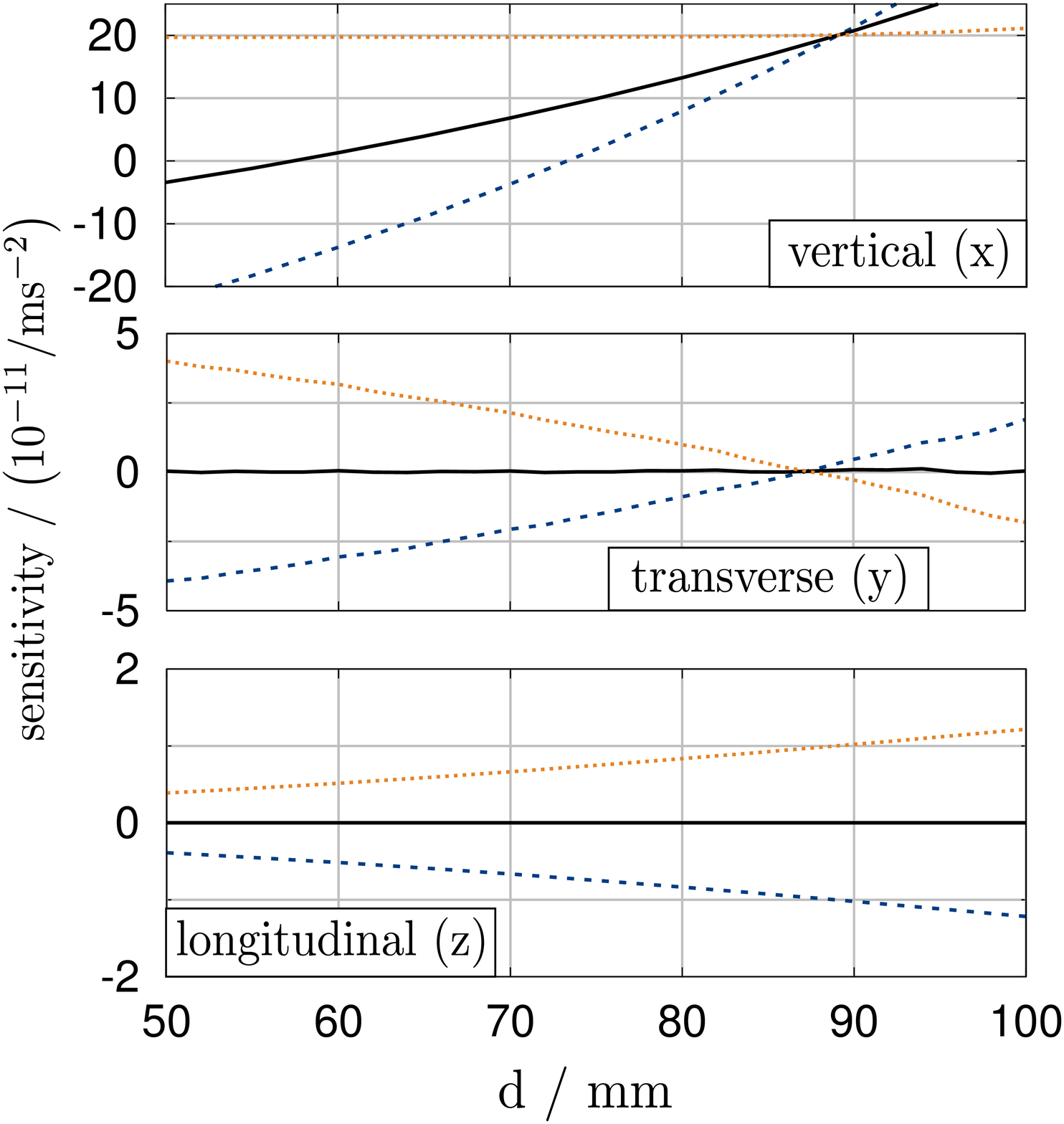}}
\caption{Calculated sensitivities as a function of the longitudinal support positions. Top to bottom, the graphs show the sensitivity to vertical, transverse and longitudinal accelerations, respectively. Solid lines correspond to the experimentally implemented placement of the optical axis. Dashed/dotted lines show the effect of $\unit[+0.45]{mm}$~/~$\unit[-0.55]{mm}$ vertical displacement, $\unit[\pm1]{mm}$ horizontal displacement, and $\unit[\pm2]{mm} / \unit[30]{cm}$ tilt, respectively.}
\label{femresults}
\end{figure}

The top graph of Fig. \ref{femresults} shows the simulated sensitivity to vertical accelerations for coinciding optical and symmetry axes (dotted), and for an optical axis that is shifted downward by $\unit[0.55]{mm}$ (solid) and $\unit[1.00]{mm}$ (dashed). We chose a displacement of $\unit[0.55]{mm}$ resulting in a zero crossing at $d=\unit[57]{mm}$. The supports are placed $\unit[5/\sqrt{2}]{mm}$ below the vertical symmetry plane. In the centre and bottom panels, the solid lines show the sensitivities to transverse and longitudinal accelerations, respectively. The dashed and dotted lines show the effect of the most critical optical axis misalignments, which is a horizontal displacement ($\unit[\pm1]{mm}$) for transverse and a vertical tilt ($\unit[\pm2]{mm}/\unit[30]{cm}$) for longitudinal vibrations. The mirror positioning tolerances are rather relaxed: a transverse displacement leads to a sensitivity of at most $\unit[4\times10^{-12}]{ms^{-2}}$ per $\unit[0.1]{mm}$, and the zero crossing of the vertical sensitivity shifts by $\unit[8]{mm}$ per $\unit[0.1]{mm}$ of vertical displacement. Due to the freely movable supports, this can easily be compensated; the slope around the zero crossing is $\unit[5\times10^{-12} /]{(ms^{-2})}$ per \unit{mm}.

\subsection{Experimental determination of the vibration sensitivity}
\label{sensitivity_measurements}
To experimentally determine the sensitivities, the laser described in section \ref{12cmcavity} is used as a reference. For these measurements, $\unit[822]{nm}$ mirrors have been contacted to the spacer using an aluminium mask to ensure correct placement of the optical axis. Both the case of the optical axis coinciding with the symmetry axis and shifted downward by $\unit[0.55]{mm}$ were investigated. We verified the accuracy of this method in the first case by a tactile determination of the mirror position and observed a displacement of $\unit[14]{\mu m}$, which is negligible when compared to the tolerances given in the previous section.

The optical setup is shown in the shaded region in figure \ref{setup}, omitting the part similar to the $\unit[12]{cm}$ cavity setup. The beam is kept in resonance with the cavity using an AOM, and the feedback (depicted as $\Delta\nu$ in the diagram) is used to deduce the relative frequency fluctuations between the cavities. A piezoelectric acceleration sensor for each dimension is placed on the vibration isolation platform and sinusoidal accelerations on the order of $\unit[5\times10^{-2}]{ms^{-2}}$ are applied for $\unit[120]{s}$ in each direction at the eigenfrequency of the platform ($\unit[0.5]{Hz}$). For evaluation, all signals are bandpass filtered and a least squares fit adjusts amplitudes of the sensitivity coefficients such that the weighted sum of the accelerations equals the frequency deviation. As expected for a rigid mount, the excitation frequency is far from any mechanical resonances and therefore no phase shift is observed. The scatter observed when repeating such a measurement without moving the cavity is below $\unit[4\times10^{-12} /]{(ms^{-2})}$.

\begin{figure}
\centerline{\includegraphics[width=.45\textwidth]{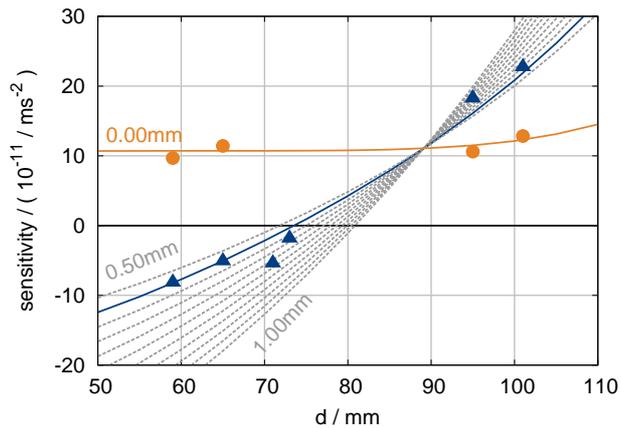}}
\caption{Experimentally observed minimum sensitivity to vertical accelerations. Circles: Optical axis coinciding with the symmetry axis. Since the mirrors tilt around the vertical symmetry plane, the length change is independent of the support position. Triangles: Optical axis $\unit[0.55]{mm}$ below the symmetry axis. The additional effect of the mirror tilt introduces a zero crossing. The lines show FEM simulations for various axis displacements, shifted by $\unit[-9\times10^{-11} /]{(ms^{-2})}$ to account for effects not included in the model; the solid lines correspond to the experimentally tested configurations. Positive accelerations point along gravity. The uncertainty of $\unit[4\times10^{-12} /]{(ms^{-2})}$ of the sensitivity measurements is smaller than the symbols.}
\label{experimental_sensitivities}
\end{figure}

Figure \ref{experimental_sensitivities} shows the observed sensitivities to vertical accelerations when varying the longitudinal support position. As expected from the simulations, there is no dependence of the support position when the optical axis coincides with the symmetry axis. In the case of a displaced optical axis, the slope agrees with the simulations, and the minimum experimental value (at $d=\unit[73]{mm}$) is $\unit[1.8\times10^{-11}/]{(ms^{-2})}$. There is an overall offset of $\unit[9\times10^{-11}/]{(ms^{-2})}$ with respect to the simulations for both mirror configurations, which means that the tilt-independent length change is two times smaller than expected. This might be due to friction at the interface with the viton spheres that reduces the transverse effect of vertical forces (the viton supports are not included in the model). As the sensitivity to accelerations in both horizontal directions only depends on the symmetry of the applied forces, both positive and negative values have been observed independently of the longitudinal support position. With careful alignment, values as low as $\unit[3.4\times10^{-12}/]{(ms^{-2})}$ have been achieved for both transverse and longitudinal accelerations.

The measurements shown here were performed with a rigid mount. Since the cavity is supported at four points, force imbalance leads to a scatter of $\pm\unit[2\times10^{-11}/]{(ms^{-2})}$ (vertical), $\pm\unit[5\times10^{-11}/]{(ms^{-2})}$ (longitudinal) and\\$\pm\unit[1\times10^{-10}/]{(ms^{-2})}$ (transverse) when the cavity is removed and realigned between measurements at the same mount position. This could be reduced by including a force balancing mechanism in the mount, e.g. allowing one of the V mounts shown in Fig. \ref{longcavity}a to tilt or translate in the $y$ direction, thus reducing the geometry to an effective three-point support. The symmetric scatter in the horizontal sensitivities (independent of the longitudinal support position) indicates that in principle, the lowest sensitivities observed so far could be obtained simultaneously.

\begin{figure}
\centerline{\includegraphics[width=.5\textwidth]{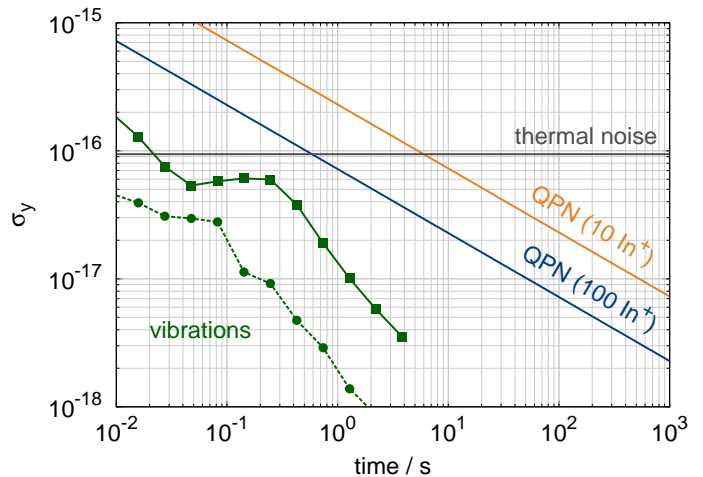}}
\caption{Expected instability contributions due to thermal noise and vibrations compared with the quantum projection noise limit for 10 and 100 In${}^+$ ions. The contributions are based on the minimum sensitivities observed simultaneously (squares: $\unit[1.8\times10^{-11}/]{(ms^{-2})}$ vertical,  $\unit[6.3\times10^{-11}/]{(ms^{-2})}$ transverse, $\unit[8.5\times10^{-11}/]{(ms^{-2})}$ longitudinal) and individually (circles: $\unit[1.8\times10^{-11}/]{(ms^{-2})}$ vertical,  $\unit[3.4\times10^{-12}/]{(ms^{-2})}$ transverse, $\unit[3.4\times10^{-12}/]{(ms^{-2})}$ longitudinal)}
\label{instability_longcavity}
\end{figure}

Figure \ref{instability_longcavity} shows the expected instability contribution due to vibrations, calculated from the spectra shown in Fig. \ref{measured_vibrations} and the lowest sensitivities observed simultaneously (squares) and individually for each dimension (circles). For comparison, the ideal case quantum projection noise limit for 10 and 100 In${}^+$ ions, calculated according to \cite{Peik2006}, is shown.

In the future, we will characterize the overall frequency instability of the indium clock laser at \unit[946]{nm} stabilized to this cavity. With the demonstrated positioning accuracy, the high-finesse mirrors can be replaced by mirrors suitable for this wavelength without limiting the achievable low vibration sensitivities.

\section{Conclusion}
\label{conclusion}
We have presented the performance of two cavities used for precision spectroscopy of  In${}^+$~/~Yb${}^+$ Coulomb crystals. A simple setup based on a cavity without vibration insensitive design provides a fractional frequency instability of $\sigma_y=4.7\times10^{-16}$ at $\unit[10]{s}$ with a linear drift of $\unit[53]{mHz/s}$. These results are in agreement with a detailed analysis of the instability contributions. Our vibration insensitive design for a  $\unit[30]{cm}$ long cavity with an expected thermal noise floor of $\sigma_y=9.7\times10^{-17}$ showed vibration sensitivities in the low $\unit[10^{-11}/]{ms^{-2}}$ range. The dependence of the vertical sensitivity to the longitudinal support position is $\unit[5\times10^{-12} /]{(ms^{-2})}$ per \unit{mm} at the zero crossing. The achieved vibration sensitivity is currently limited by unequal forces in the spacer support and could be further reduced by implementing a force-balancing mount, e.g. by allowing one of the supporting V blocks to tilt around the longitudinal axis (see Fig. \ref{longcavity}a). This would be of use in setups with a lower thermal noise floor, e.g. when using coatings with lower mechanical losses \cite{Cole2013}. However, with the observed sensitivities and expected thermal noise, this cavity already provides a sufficiently stable short-term reference to operate an optical frequency standard based on 100 In${}^+$ ions at its quantum projection noise limit.

\begin{acknowledgement}
The authors would like to thank C. Grebing for conducting the frequency comb based measurements, M. Okhapkin and C. Tamm  for fruitful discussions, P.O. Schmidt for comments on the manuscript, and T. Legero  for stimulating discussions and optically contacting the high-reflectivity mirrors. This work was supported by DFG through QUEST and DFG/RFBR (Grant No. 10-02-91335).
\end{acknowledgement}


\begin{thebibliography}{10}
\providecommand{\url}[1]{{#1}}
\providecommand{\urlprefix}{URL }
\expandafter\ifx\csname urlstyle\endcsname\relax
  \providecommand{\doi}[1]{DOI \discretionary{}{}{}#1}\else
  \providecommand{\doi}{DOI \discretionary{}{}{}\begingroup
  \urlstyle{rm}\Url}\fi

\bibitem{Chou2010}
C. {Chou}, D. {Hume}, J. {Koelemeij}, D. {Wineland}, T. {Rosenband}, Phys. Rev.
  Lett. \textbf{104}, 070802 (2010)

\bibitem{Itano1993}
W. {Itano}, J. {Bergquist}, J. {Bollinger}, J. {Gilligan}, D. {Heinzen},
  F. {Moore}, M. {Raizen}, D. {Wineland}, Phys. Rev. A \textbf{47}, 3554 (1993)

\bibitem{Herschbach2011}
N. Herschbach, K. Pyka, J. Keller, T. {Mehlst\"aubler}, Appl. Phys. B
  \textbf{107}, 891 (2012)

\bibitem{Pyka2012}
K. Pyka, N. Herschbach, J. Keller, T. {Mehlst\"aubler}, Appl. Phys. B  (2013).
\newblock \doi{10.1007/s00340-013-5580-5}

\bibitem{Peik2006}
E. {Peik}, T. {Schneider}, C. {Tamm}, J. Phys. B \textbf{39}, 145 (2006)

\bibitem{Young1999}
B. {Young}, F. {Cruz}, W. {Itano}, J. {Bergquist}, Phys. Rev. Lett.
  \textbf{82}, 3799 (1999)

\bibitem{Numata2004}
K. {Numata}, A. {Kemery}, J. {Camp}, Phys. Rev. Lett. \textbf{93}, 250602
  (2004)

\bibitem{Nazarova2006}
T. {Nazarova}, F. {Riehle}, U. {Sterr}, Appl. Phys. B \textbf{83}, 531 (2006)

\bibitem{Ludlow2007}
A. {Ludlow}, X. {Huang}, M. {Notcutt}, T. {Zanon-Willette}, S. {Foreman},
  M. {Boyd}, S. {Blatt}, J. {Ye}, Opt. Lett. \textbf{32}, 641 (2007)

\bibitem{Webster2008}
S. {Webster}, M. {Oxborrow}, S. {Pugla}, J. {Millo}, P. {Gill}, Phys. Rev. A
  \textbf{77}, 033847 (2008)

\bibitem{Millo2009}
J. {Millo}, D. {Magalh{\~a}es}, C. {Mandache}, Y. {Le Coq}, E. {English},
  P. {Westergaard}, J. {Lodewyck}, S. {Bize}, P. {Lemonde}, G. {Santarelli},
  Phys. Rev. A \textbf{79}, 053829 (2009)

\bibitem{Dawkins2010}
S. {Dawkins}, R. {Chicireanu}, M. {Petersen}, J. {Millo}, D. {Magalh{\~a}es},
  C. {Mandache}, Y. {Le Coq}, S. {Bize}, Appl. Phys. B \textbf{99}, 41 (2010)

\bibitem{Kessler2012}
T. {Kessler}, C. {Hagemann}, C. {Grebing}, T. {Legero}, U. {Sterr},
  F. {Riehle}, M. {Martin}, L. {Chen}, J. {Ye}, Nat. Photonics \textbf{6}, 687
  (2012)

\bibitem{DAmbrosio2003}
E. {D'Ambrosio}, Phys. Rev. D \textbf{67}, 102004 (2003)

\bibitem{Mours2006}
B. {Mours}, E. {Tournefier}, J.Y. {Vinet}, Classical Quant. Grav. \textbf{23},
  5777 (2006)

\bibitem{Amairi2012}
S. Amairi, T. Legero, T. Kessler, U. Sterr, J. W\"ubbena, O. Mandel,
  P. Schmidt, Appl. Phys. B  (2013).
\newblock \doi{10.1007/s00340-013-5464-8}

\bibitem{Cole2013}
G. Cole, W. Zhang, M. Martin, J. Ye, M. Aspelmeyer, Nat. Photonics \textbf{7}, 644 (2013)

\bibitem{Jiang2011}
Y. Jiang, A. Ludlow, N. Lemke, R. Fox, J. Sherman, L.S. Ma, C. Oates, Nat.
  Photonics \textbf{5}, 158 (2011)

\bibitem{Nicholson2012}
T. {Nicholson}, M. {Martin}, J. {Williams}, B. {Bloom}, M. {Bishof},
  M. {Swallows}, S. {Campbell}, J. {Ye}, Phys. Rev. Le \textbf{109}, 230801
  (2012)

\bibitem{Haefner}
S. {H\"afner}, U. Sterr.
\newblock priv. comm.

\bibitem{ULEdatasheet}
{Corning Inc.}
\newblock {ULE$\textsuperscript{\textregistered}$ Corning Code 7973 data sheet}

\bibitem{Legero2010}
T. {Legero}, T. {Kessler}, U. {Sterr}, J. Opt. Soc. Am. B \textbf{27}, 914
  (2010)

\bibitem{Schawlow1958}
A. {Schawlow}, C. {Townes}, Phys. Rev. \textbf{112}, 1940 (1958)

\bibitem{Zhao2010}
Y. Zhao, J. Zhang, J. Stuhler, G. Schuricht, F. Lison, Z. Lu, L. Wang, Opt.
  Commun. \textbf{283}, 4696  (2010)

\bibitem{Littman1978}
M. {Littman}, H. {Metcalf}, Appl. Optics \textbf{17}, 2224 (1978)

\bibitem{Drever1983}
R. {Drever}, J. {Hall}, F. {Kowalski}, J. {Hough}, G. {Ford}, A. {Munley},
  H. {Ward}, Appl. Phys. B \textbf{31}, 97 (1983)

\bibitem{Ma1994}
L.S. {Ma}, P. {Jungner}, J. {Ye}, J. {Hall}, Opt. Lett. \textbf{19}, 1777
  (1994)

\bibitem{Gray1974}
J. Gray, D. Allan, Proc. 28th Frequency Control Symposium p. 243 (1974)

\bibitem{Telle2002}
H. {Telle}, B. {Lipphardt}, J. {Stenger}, Appl. Phys. B \textbf{74}, 1 (2002)

\end{thebibliography}
\end{document}